\documentclass[superscriptaddress,onecolumn]{revtex4}
\usepackage{amsmath}
\usepackage{amssymb}
\usepackage{graphicx}
\usepackage{hyperref}
\usepackage{amsfonts}
\usepackage{bbm}
\usepackage{doi}
\usepackage{todonotes}

\begin{document}
\title{Intrinsic Geometry and Director Reconstruction for Three-Dimensional Liquid Crystals}
\author{Joseph Pollard}
\affiliation{Mathematics Institute, Zeeman Building, University of Warwick, Coventry, CV4 7AL, United Kingdom.}
\author{Gareth P. Alexander}
\email{G.P.Alexander@warwick.ac.uk}
\affiliation{Department of Physics and Centre for Complexity Science, University of Warwick, Coventry, CV4 7AL, United Kingdom.}

\begin{abstract}
We give a description of the intrinsic geometry of elastic distortions in three-dimensional nematic liquid crystals and establish necessary and sufficient conditions for a set of functions to represent these distortions by describing how they couple to the curvature tensor. We demonstrate that, in contrast to the situation in two dimensions, the first-order gradients of the director alone are not sufficient for full reconstruction of the director field from its intrinsic geometry: it is necessary to provide additional information about the second-order director gradients. We describe several different methods by which the director field may be reconstructed from its intrinsic geometry. Finally, we discuss the coupling between individual distortions and curvature from the perspective of Lie algebras and groups and describe homogeneous spaces on which pure modes of distortion can be realised. 
\end{abstract}
\date{\today}
\maketitle

The geometric characterisation of liquid crystals textures has been fundamental to their understanding. A classic example of the insights of geometric methodology is the description of the cholesteric blue phases as the result of the geometric frustration of trying to realise in flat space the perfect double twist texture that exists on $S^3$~\cite{sethna1983}. Recently, Sadoc {\sl et al.}~\cite{sadoc20} have extended this to give a more general description of three-dimensional liquid crystal textures frustrated by spatial curvature and to identify the geometries in which pure geometric distortions can be perfectly realised. However, geometric methods have broader utility than the identification of `pure' geometric states. A general analysis of the liquid crystal director can be developed through a geometric decomposition of its gradients~\cite{machon16}. This allows for geometrically distinguished directions to be identified, for instance the pitch axis in cholesterics~\cite{efrati2014,beller2014}, and connects the defects in these directions to the topology of the texture. Selinger~\cite{selinger18} has emphasised how this approach can give fresh perspectives on many standard liquid crystal textures and on the Frank free energy itself. Indeed, Frank's original derivation~\cite{frank58} of the free energy of liquid crystals is founded upon the geometry of their elastic distortions; splay, twist, bend and saddle-splay. Geometric methods continue to furnish insights into liquid crystals phases: the geometry of bend distortions gives a description of defects and textures in the twist-bend nematic phase of bent-core molecules~\cite{binysh2020}, while a geometric anaylsis of the packing of diabolos offers a similar insight into the formation of structures in the splay-twist phase~\cite{chaturvedi2020,kamien2020}, and a further recent application has been to the analysis of defect lines in active liquid crystals~\cite{binysh2020_2,long2020}.

The geometric decomposition of the director gradients affords for an intrinsic description of liquid crystal textures, analogous to the intrinsic geometry of curves or surfaces.
For example, for surfaces the intrinsic geometry is the mean and Gaussian curvature, and more generally the first and second fundamental forms.
In the case of both curves and surfaces, in addition to the intrinsic characterisation there is also the ability to reconstruct the curve or surface from its intrinsic geometry, albeit subject to certain compatibility conditions. One may ask if a similar geometric reconstruction can be formulated for the liquid crystal director. This problem was resolved positively for two-dimensional director fields by Niv \& Efrati~\cite{niv18} and involves a compatibility between the intrinsic geometry and the curvature of space. Here, we address the problem for three-dimensional director fields. Our approach is based on Cartan's method of moving frames~\cite{cartan04,cartan37,cartan45,olver} and again directly connects the distortions of the liquid crystal director to expressions for the curvature of space, which provide compatibility equations and lead to a reconstruction formula.

This paper is organised as follows. In Sec. \ref{sec:geometric_reconstruction_2d} we review the intrinsic description of a two-dimensional director as presented by Niv \& Efrati~\cite{niv18}, which is reinterpreted in the language of moving frames in Sec. \ref{sec:frames}. A direct extension to three-dimensions is given in Sec. \ref{sec:3dgeometry}. The greater degree of freedom in three-dimensions poses a challenge for reconstruction of the director, as the splay, twist, and bend alone are not sufficient. In Sec. \ref{sec:geometric_reconstruction_3d} we solve the reconstruction problem. A direct generalisation of the method proposed by Niv \& Efrati~\cite{niv18} is possible, and we also describe other approaches that require specifying different levels of information about the director gradients. In Sec. \ref{sec:lie_groups}, we describe the connection between our description of the director and the director's symmetry group. This offers a fresh perspective on the geometry of pure distortions discussed by Virga~\cite{virga19}, and more recently by Sadoc {\sl et al.}~\cite{sadoc20}, and provides a framework for extending this analysis to directors whose distortions are not uniform but still possess a degree of regularity. The paper concludes in Sec. \ref{sec:discussion} with a discussion.

\section{Geometric Compatibility and Reconstruction in Two Dimensions} \label{sec:geometric_reconstruction_2d}

To motivate our constructions, we first review the compatibility condition and reconstruction formula for a director in two dimensions, given by Niv \& Efrati~\cite{niv18}. Given any director ${\bf n}$ there is a unique unit vector ${\bf n}_{\perp}$ orthogonal to it such that the pair ${\bf n},{\bf n}_{\perp}$ form a right-handed basis for the tangent space. Let $\eta, \eta_\perp$ denote the differential 1-forms dual to ${\bf n}, {\bf n}_\perp$. The derivatives of ${\bf n}$ decompose as
\begin{equation}
  \nabla {\bf n} = \bigl(\kappa \eta + s\eta_{\perp}\bigr) \otimes {\bf n}_\perp,
\end{equation}
where $\kappa$ is the curvature of the integral curves of ${\bf n}$ (and magnitude of the bend ${\bf b} = \nabla_{{\bf n}}{\bf n}$) and $s = \nabla \cdot {\bf n}$ is the splay. We read off the connection 1-form $\omega = \kappa \eta + s\eta_\perp$. The curvature 2-form is $\Omega = d\omega = d\kappa \wedge \eta + \kappa d\eta + ds \wedge \eta_\perp + s d\eta_\perp$, and the Gaussian curvature of the two-dimensional surface is $K = -\Omega({\bf n}, {\bf n}_\perp)$. Using the formula $\Omega(X, Y) = X(\omega(Y)) - Y(\omega(X)) - \omega([X, Y])$ for the curvature form, we write this as
\begin{equation}
  K = \nabla_{{\bf n}_{\perp}}\kappa - \nabla_{{\bf n}} s + \kappa \eta\bigl([{\bf n}, {\bf n}_{\perp}]\bigr) + s \eta_{\perp}\bigl([{\bf n}, {\bf n}_{\perp}]\bigr) ,
\end{equation}
and since the connection is torsion free we have $[{\bf n}, {\bf n}_\perp] = \nabla_{{\bf n}} {\bf n}_\perp - \nabla_{{\bf n}_\perp} {\bf n} = -\kappa {\bf n} - s{\bf n}_\perp$, which leads to the geometric compatibility equation~\cite{niv18}
\begin{equation} \label{eq:compatibility_2d}
  K = -s^2 - \kappa^2 - \nabla_{{\bf n}} s + \nabla_{{\bf n}_\perp} \kappa .
\end{equation}

The reconstruction problem in two dimensions is to find a director field given (generic) splay and bend functions, $s$ and $\kappa$. The geometric compatibility condition~\eqref{eq:compatibility_2d} is key to solving it. We let $J$ denote the almost complex structure on the tangent space and then, by noting that ${\bf n}_{\perp}\cdot\nabla\kappa = (J{\bf n})\cdot\nabla\kappa = - {\bf n} \cdot J\nabla\kappa$, rewrite the compatibility condition~\eqref{eq:compatibility_2d} as
\begin{equation}
{\bf n} \cdot \bigl( \nabla s + J \nabla \kappa \bigr) = -K - s^2 - \kappa^2 .
\label{eq:compatibility_2d_2}
\end{equation}
The reconstruction problem can now be solved as follows: Using the given splay and bend functions we can define the orthonormal frame
\begin{align}
& {\bf e}_1 = \frac{\nabla s + J \nabla \kappa}{|\nabla s + J \nabla \kappa|} , && {\bf e}_2 = J {\bf e}_1 ,
\end{align}
and the compatibility condition gives the component of the director along ${\bf e}_1$; the component along ${\bf e}_2$ follows from normalisation, $|{\bf n}|=1$. Explicitly we have the reconstruction formula
\begin{equation}
  {\bf n} = - \frac{s^2+\kappa^2+K}{|\nabla s + J \nabla \kappa|} \,{\bf e}_1 \pm \sqrt{1 - \frac{(s^2+\kappa^2+K)^2}{|\nabla s + J \nabla \kappa|^2}} \,{\bf e}_2.
  \label{eq:reconstruction_2d}
\end{equation}
The sign choice is not arbitrary: only one of the branches yields the correct director~\cite{niv18}. To resolve the ambiguity we compute the splay and bend of the director and choose the sign so that they agree with $s$ and $\kappa {\bf n}_\perp$.
It is interesting that the reconstruction formula is purely algebraic. This contrasts with the reconstruction of space curves and surfaces in $\mathbb{R}^3$, both of which require solving a differential equation.

The unresolved part of the reconstruction is the identification of the allowed set of functions that can represent the splay and bend of a director field.
The geometric compatibility equation~\eqref{eq:compatibility_2d_2} implies that the functions $s$ and $\kappa$ must satisfy the necessary condition
\begin{equation} \label{eq:necessary_condition}
  | \nabla s + J \nabla \kappa | \geq \bigl| s^2 + \kappa^2 + K \bigr| .
\end{equation}
For example, on a flat surface ($K=0$) both $s$ and $\kappa$ must vanish at any points where $\nabla s + J \nabla \kappa = 0$. The vanishing of this combination of gradients is equivalent to the Cauchy--Riemann equations for $s+i\kappa$; it follows that $s$ and $\kappa$ cannot be the real and imaginary parts of a holomorphic function unless they are both constant and $s^2+\kappa^2 = -K$. However, the constraint represented by the necessary condition~\eqref{eq:necessary_condition} is stronger than this and we do not currently know what the set of allowed splay and bend functions is.

\begin{figure*}[tb]
\centering
\includegraphics[width=\textwidth]{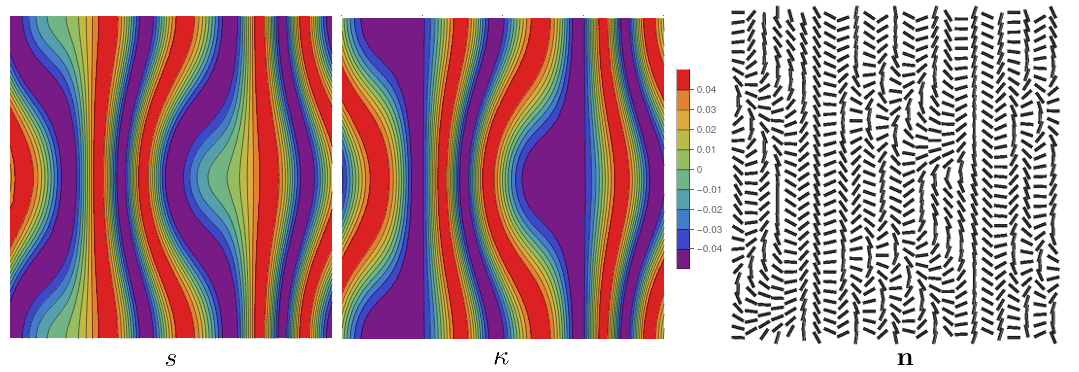}
\caption{Illustration of geometric compatibility in 2D. Specify a pair of functions, the splay $s$ and the curvature $\kappa$. The geometric compatibility equation allows us to construct a director ${\bf n}$. Here we have $s = \tfrac{1}{20}\cos \theta$ and $\kappa = \tfrac{1}{20}\sin \theta$, where $\theta = \cos(\pi x) + \cos\left(\tfrac{\pi}{2}(x-\sqrt{3}y)\right) + \cos\left(\tfrac{\pi}{2}(x+\sqrt{3}y)\right)+2\pi x$.}
\label{fig:reconstruction2d}
\end{figure*}

For an example, we consider the case of a flat space, $K=0$, with $s=a\cos \theta, \kappa = a\sin \theta$, for a function $\theta$ and a constant scale factor $a$. The condition \eqref{eq:necessary_condition} then simplifies to $|a\nabla \theta|^2 \geq 1$, which may be easily satisfied.
For an example, we show a director with a broken hexagonal symmetry, given by $\theta = \cos(\pi x) + \cos\left(\tfrac{\pi}{2}(x-\sqrt{3}y)\right) + \cos\left(\tfrac{\pi}{2}(x+\sqrt{3}y)\right)+2\pi x$ and scale factor $a=0.05$. The first three terms impose the hexagonal symmetry, while the final term ensures that \eqref{eq:necessary_condition} is satisfied. The functions $s, \kappa$ that result from this choice are shown in the unit cell in Fig.~\ref{fig:reconstruction2d}, along with the reconstructed director ${\bf n}$.

\section{Frames, Connections and Structure Functions}
\label{sec:frames}

As a precursor to considering the corresponding compatibility and reconstruction problems in three dimensions, we describe how the two-dimensional case fits into a more general picture. A geometric reconstruction of the director from its gradients is possible in two dimensions because any director field immediately extends to an orthonormal frame, $\{{\bf e}_1={\bf n}, \,{\bf e}_2={\bf n}_{\perp}\}$, and it is a classical result that a frame is completely determined by its {\it structure functions} $c_{ij}^k$ giving the Lie brackets for the frame~\cite{cartan37,cartan45,olver}
\begin{equation}
[{\bf e}_i , {\bf e}_j] = \sum_k c_{ij}^k \,{\bf e}_k .
\label{eq:structurefunctions}
\end{equation}
When the connection is torsion free, the structure functions are equivalent to the components of the connection $\omega_{ij}^k$ describing the gradients of the frame
\begin{equation}
\nabla_{{\bf e}_i} {\bf e}_j = \sum_k \omega_{ij}^k \,{\bf e}_k ,
\end{equation}
with the relationship being $c_{ij}^k = \omega_{ij}^k - \omega_{ji}^k$, or equivalently
\begin{equation} \label{eq:connection}
  \omega_{ij}^k = \frac{1}{2} \bigl( c_{ij}^k + c_{ki}^j + c_{kj}^i \bigr) .
\end{equation}
In two dimensions there is only one Lie bracket, $[{\bf e}_1, {\bf e}_2]$ and two structure functions, $c_{12}^1$ and $c_{12}^2$, which are equivalent to the splay and bend
\begin{align}
& c_{12}^1 = - \kappa , && c_{12}^2 = - s .
\end{align}
Said conversely, the splay and bend determine all of the structure functions and consequently are sufficient to reconstruct the frame.
In three dimensions there are nine structure functions (equivalently components of the connection) so that more information than just the basic director distortions of splay, twist and bend is needed to specify them all and facilitate a geometric reconstruction.

\section{Geometry of Three-Dimensional Director Fields}
\label{sec:3dgeometry}

The director splits directions in space (at each point) into those parallel to it and those perpendicular, which we call the orthogonal plane $\xi$, and we decompose the director gradients with respect to this splitting. The parallel gradients, $\nabla_{\parallel}{\bf n}$, contain the bend, while those along perpendicular directions, $\nabla_{\perp}{\bf n}$ correspond to the shape operator of the director field; these decompose further into pieces that transform independently under the local $SO(2)$ symmetry of rotations about the director~\cite{machon16,alexander18} as
\begin{equation}
  \begin{split}
    \nabla_{\perp}{\bf n} & = \frac{s}{2} \bigl( {e}^1{\bf e}_1 + {e}^2{\bf e}_2 \bigr) + \frac{q}{2} \bigl( {e}^1{\bf e}_2 - {e}^2{\bf e}_1 \bigr) \\
    & \quad + \Delta_1 \bigl( {e}^1{\bf e}_1 - {e}^2{\bf e}_2 \bigr) + \Delta_2 \bigl( {e}^1{\bf e}_2 + {e}^2{\bf e}_1 \bigr) ,
  \end{split}
\label{eq:perp_gradients}
\end{equation}
where ${\bf e}_1$, ${\bf e}_2$ are a basis for the orthogonal planes, $e^1, e^2$ are their dual 1-forms, $s=\nabla\cdot{\bf n}$ is the splay, $q={\bf n}\cdot\nabla\times{\bf n}$ is the twist, and $\Delta_1$, $\Delta_2$ are the components of the spin 2 deviatoric part of the orthogonal gradients.
Although the deviatoric part ($\Delta$) has intrinsic geometrical, and topological, significance, it does not (usually) contribute a term in the Frank free energy, alongside splay, twist and bend, because the (flat space) identity
\begin{equation} \label{eq:saddle-splay}
\nabla \cdot \bigl[ ({\bf n}\cdot\nabla) {\bf n} - (\nabla\cdot{\bf n}) {\bf n} \bigr] + \frac{s^2}{2} + \frac{q^2}{2} - |\Delta|^2 = 0 ,
\end{equation}
allows it to be eliminated modulo a total divergence -- the saddle-splay~\cite{machon16,selinger18}. This identity represents one of three curvature compatability conditions, as we describe below.

The director by itself does not differentiate any directions in the orthogonal plane to give an entire frame field, however for a generic director its gradients do, allowing for a geometric construction of a frame. In fact there are several possible choices; we present our subsequent analysis only for the Frenet-Serret frame of the director field. This is the frame associated to the integral curves of the director and defined by the bend: the bend ${\bf b} = \nabla_{{\bf n}} {\bf n} = - {\bf n} \times (\nabla\times{\bf n})$ is a vector that is everywhere perpendicular to the director, ${\bf b}\cdot{\bf n}=0$; we write ${\bf b} = \kappa \,{\bf e}_1$, where $\kappa$ is the (magnitude of the) curvature of the director integral curves and ${\bf e}_1$ is their Frenet-Serret normal~\cite{binysh2020}. We then define ${\bf e}_2 = {\bf n} \times {\bf e}_1$ to complete the Frenet-Serret frame associated to the director field.

The director gradients determine two of the connection 1-forms
\begin{align}
  \omega_{3}^1 & = \bigl( \nabla {\bf n} \bigr) \cdot {\bf e}_1 = \biggl( \frac{s}{2} + \Delta_1 \biggr) e^1 + \biggl( \Delta_2 - \frac{q}{2} \biggr) e^2 + \kappa \,\eta , \\
  \omega_{3}^2 & = \bigl( \nabla {\bf n} \bigr) \cdot {\bf e}_2 = \biggl( \Delta_2 + \frac{q}{2} \biggr) e^1 + \biggl( \frac{s}{2} - \Delta_1 \biggr) e^2 ,
\end{align}
where $\eta$ is the 1-form dual to the director.
The third connection 1-form is the connection of the Frenet-Serret frame for the planes orthogonal to the director~\cite{binysh2020}
\begin{equation}
  \omega_{1}^2 = \bigl( \nabla {\bf e}_1 \bigr) \cdot {\bf e}_2 = \omega_{11}^2 \,e^1 + \omega_{21}^2 \,e^2 + \tau \,\eta .
\end{equation}
Here, $\tau$ is the torsion of the director integral curves, however, the components $\omega_{11}^2$ and $\omega_{21}^2$ do not, as far as we know, have their own terminology and interpretation in terms of the director field. Explicitly, $\omega_{11}^2 = (\nabla_{{\bf e}_1}{\bf e}_1)\cdot{\bf e}_2$ is the bend of the Frenet-Serret normal within the orthogonal planes and analogously $\omega_{21}^2 = (\nabla_{{\bf e}_2}{\bf e}_1)\cdot{\bf e}_2$ is the splay of the Frenet-Serret normal within the orthogonal planes.
Nonetheless, we emphasise that these two remaining components of the connection are just as directly determined from any generic director field as the other components. All nine are required in order to reconstruct the (Frenet-Serret frame for the) director.

The nine components of the connection 1-forms can equivalently be expressed in terms of the structure functions of the frame using Eq. (\ref{eq:connection})
\begin{equation}
\begin{split}
& c_{12}^1 = - \omega_{11}^2 , \quad c_{13}^1 = \frac{s}{2} + \Delta_1 , \qquad c_{23}^1 = \Delta_2 - \frac{q}{2} + \tau , \\
& c_{12}^2 = - \omega_{21}^2 , \quad c_{13}^2 = \Delta_2 + \frac{q}{2} - \tau , \quad c_{23}^2 = \frac{s}{2} - \Delta_1 , \\
& c_{12}^3 = - q , \qquad c_{13}^3 = \kappa , \quad c_{23}^3 = 0 .
\end{split}
\end{equation}

The reconstruction is facilitated by expressions for the curvature in terms of the components of the connection, or structure functions, in a similar fashion to the two dimensional case. The curvature 2-form has components
\begin{equation}
 \Omega_j^k = d\omega_j^k + \sum_i \omega^k_i \wedge \omega_j^i = \frac{1}{2}\sum_{il} R_{ilj}^k e^i \wedge e^l,
\end{equation}
where $R_{ilj}^k$ are the components of the Riemann tensor. There are six independent components of this tensor, and in Euclidean space they all vanish, providing the compatibility conditions. In terms of the connection 1-form, the components of Riemann are
\begin{equation} \label{eq:Riemann}
  R_{ijk}^l = \nabla_{{\bf e}_i} \omega_{jk}^l - \nabla_{{\bf e}_j} \omega_{ik}^l - \sum_m \bigl[ (\omega_{ij}^m - \omega_{ji}^m) \omega_{mk}^l - \omega_{jk}^m \omega_{im}^l + \omega_{ik}^m \omega_{jm}^l \bigr] ,
\end{equation}
and the six independent components are given explicitly by
\begin{align}
  R_{122}^1 & = \nabla_{{\bf e}_2} \omega_{11}^2 - \nabla_{{\bf e}_1} \omega_{21}^2 - \bigl( \omega_{11}^2 \bigr)^2 - \bigl( \omega_{21}^2 \bigr)^2 - \frac{s^2}{4} - \frac{q^2}{4} + \frac{1}{2} |\Delta|^2 - q \tau , \label{eq:K12} \\
  R_{133}^1 & = \nabla_{{\bf e}_1} \kappa - \nabla_{{\bf n}} \biggl( \frac{s}{2} + \Delta_1 \biggr) - \biggl( \frac{s}{2} + \Delta_1 \biggr)^2 - \Delta_2^2 + \frac{q^2}{4} - \kappa^2 + 2 \tau \Delta_2 , \\
  R_{233}^2 & = \nabla_{{\bf n}} \biggl( \Delta_1 - \frac{s}{2} \biggr) - \biggl( \frac{s}{2} - \Delta_1 \biggr)^2 - \Delta_2^2 + \frac{q^2}{4} + \kappa \omega_{21}^2 - 2 \tau \Delta_2 , \\
  R_{123}^1 & = \nabla_{{\bf e}_1} \biggl( \Delta_2 - \frac{q}{2} \biggr) - \nabla_{{\bf e}_2} \biggl( \frac{s}{2} + \Delta_1 \biggr) + 2 \omega_{11}^2 \Delta_1 + 2 \omega_{21}^2 \Delta_2 + \kappa q , \\
  R_{123}^2 & = \nabla_{{\bf e}_1} \biggl( \frac{s}{2} - \Delta_1 \biggr) - \nabla_{{\bf e}_2} \biggl( \Delta_2 + \frac{q}{2} \biggr) + 2 \omega_{11}^2 \Delta_2 - 2 \omega_{21}^2 \Delta_1 , \\
  R_{133}^2 & = - \nabla_{{\bf n}} \biggl( \Delta_2 + \frac{q}{2} \biggr) - s \biggl( \Delta_2 + \frac{q}{2} \biggr) - 2 \tau \Delta_1 + \kappa \omega_{11}^2 .
\end{align}
The first three of these correspond to the sectional curvatures $K_{ij} = - \Omega_i^j({\bf e}_i,{\bf e}_j)=-R_{iji}^j$ of the `Frenet-Serret planes'. Further insight is gained by adding the second and third to give
\begin{equation} \label{eq:saddle_splay}
\begin{aligned}
K_{13} + K_{23} & = \nabla_{{\bf e}_1} \kappa - \nabla_{{\bf n}} s - \frac{s^2}{2} - 2 \bigl( \Delta_1^2 + \Delta_2^2 \bigr) + \frac{q^2}{2} - \kappa^2 + \kappa \omega^2_{21} , \\
& = \nabla \cdot \bigl[ ({\bf n}\cdot\nabla) {\bf n} - (\nabla\cdot{\bf n}) {\bf n} \bigr] + \frac{s^2}{2} + \frac{q^2}{2} - |\Delta|^2 .
\end{aligned}
\end{equation}
In flat space, where the curvatures all vanish, this reduces to the familiar relation \eqref{eq:saddle-splay}. In a general curved space the sum of sectional curvatures $K_{13}+K_{23}=R_{33}$ is the (33)-component of the Ricci tensor, {\sl i.e.} it is $\textrm{Ric}({\bf n}, {\bf n})$.

We remark that the sectional curvature $K_{23}=R_{233}^2$ determines the component of the director along the direction $\nabla(s/2 - \Delta_1)$. As in two dimensions this then gives the constraint
\begin{equation} \label{eq:k23}
\biggl| \nabla \biggl( \frac{s}{2} - \Delta_1 \biggr) \biggr|^2 \geq \biggl[ K_{23} + \biggl( \frac{s}{2} - \Delta_1 \biggr)^2 + \Delta_2^2 - \frac{q^2}{4} + 2\Delta_2 \tau - \kappa \omega^2_{21} \biggr]^2 ,
\end{equation}
on the director distortions and their gradients. In exactly the same way the component $R_{133}^2$ leads to the gradient constraint
\begin{equation}
\biggl| \nabla \biggl( \Delta_2 + \frac{q}{2} \biggr) \biggr|^2 \geq \biggl[ R_{133}^2 + s \biggl( \Delta_2 + \frac{q}{2} \biggr) + 2 \tau \Delta_1 - \kappa \omega_{11}^2 \biggr]^2 .
\end{equation}
As in two dimensions, we do not currently know the full restriction that these relations imply for the set of allowable director distortions.

In three dimensions these curvature compatibility conditions are not the only constraints on the director. There are three additional constraints, which we call the algebraic compatibility conditions. They can be expressed most easily in terms of the dual coframe $e^1, e^2, e^3 = \eta$ as we now describe. The derivative of each element in the dual coframe can be written
\begin{equation} \label{eq:coframe_structure_functions}
  de^k = \sum_{i < j} -c_{ij}^k e^i \wedge e^j,
\end{equation}
where as before $c_{ij}^k$ are the structure functions. The algebraic compatibility conditions derive from the fact that the 2-forms $de^k$ are closed,
\begin{equation} \label{eq:algebraic_compatibility_condition}
  \begin{aligned}
  0 = d^2e^k = &-c_{12}^k\left(c_{13}^1+c_{23}^2 \right) + c_{13}^k\left(c_{12}^1-c_{23}^3 \right) \\
                &+ c_{23}^k\left(c_{12}^2+c_{13}^3 \right) - \nabla_{{\bf e}_1} c_{23}^k + \nabla_{{\bf e}_2} c_{13}^k - \nabla_{{\bf e}_3} c_{12}^k,
  \end{aligned}
\end{equation}
for $k=1,2,3$. Note that these conditions do not exist in two dimensions as $de^k$ is then a top form and its derivative vanishes for dimensional reasons alone. These constraints derive from the algebraic properties of the exterior derivative, or equivalently the Liebniz identity for the Lie bracket, and cannot be alleviated by curvature.
Written out in terms of the director distortions, the three algebraic compatibility conditions are
\begin{gather}
\begin{split}
& \nabla_{{\bf e}_1} \biggl( \Delta_2 - \frac{q}{2} + \tau \biggr) - \nabla_{{\bf e}_2} \biggl( \frac{s}{2} + \Delta_1 \biggr) - \nabla_{{\bf n}} \omega^2_{11} = \\
& \quad \omega^2_{11} \biggl( \frac{s}{2} - \Delta_1 \biggr) + \bigl( \kappa - \omega^2_{21} \bigr) \biggl( \Delta_2 - \frac{q}{2} + \tau \biggr) ,
\end{split} \\
\begin{split}
& \nabla_{{\bf e}_1} \biggl( \frac{s}{2} - \Delta_1 \biggr) + \nabla_{{\bf e}_2} \biggl( \tau - \Delta_2 - \frac{q}{2} \biggr) - \nabla_{{\bf n}} \omega^2_{21} = \\
& \quad s \omega^2_{21} + \omega^2_{11} \biggl( \tau - \Delta_2 - \frac{q}{2} \biggr) + \bigl( \kappa - \omega^2_{21} \bigr) \biggl( \frac{s}{2} - \Delta_1 \biggr) ,
\end{split} \\
\nabla_{{\bf e}_2} \kappa + \nabla_{{\bf n}} q = \kappa \omega^2_{11} - qs .
\end{gather}
The last of these can be written equivalently as
\begin{equation} \label{eq:projector_q}
{\bf n} \cdot \nabla q - {\bf n} \cdot \nabla \times {\bf b} + qs = 0 ,
\end{equation}
and even expressed purely in terms of the director field as
\begin{equation}
\nabla \cdot \bigl[ {\bf n} \bigl( {\bf n} \cdot \nabla \times {\bf n} \bigr) \bigr] = {\bf n} \cdot \nabla \times \bigl[ ({\bf n} \cdot \nabla) {\bf n} \bigr] ,
\end{equation}
which can be verified by a short, direct calculation. We are unaware of similarly simple interpretations of the other constraints.

We identify one additional geometric quantity of interest, associated to the planes $\xi$ orthogonal to the director. By restricting the Euclidean metric to this bundle we obtain a Riemannian metric $g$ on $\xi$; if $I$ denotes the Euclidean metric, then $g$ is simply the orthogonal projector $I - {\bf n} \otimes {\bf n}$. When the twist of ${\bf n}$ vanishes the director is the normal to a family of surfaces and the metric $g$ is the induced metric on these surfaces, or first fundamental form. Even when the twist does not vanish we may still regard $g$ as being the `first fundamental form' of the director, in the same way that (the symmetric part of) $\nabla_\perp {\bf n}$ plays the role of the second fundamental form.

The curvature $K_\xi$ of $g$ may be computed via a basis of orthonormal vector fields for $\xi$ in the same fashion as the curvature of a surface, using the formulae of \S\ref{sec:geometric_reconstruction_2d}. In terms of the connection 1-forms of an orthonormal frame ${\bf e}_1, {\bf e}_2$ for $\xi$, we have $K_\xi = -(\omega_{11}^2)^2-(\omega_{21}^2)^2 - \nabla_{{\bf e}_1} \omega_{21}^2 + \nabla_{{\bf e}_2} \omega_{11}^2$. This quantity is independent of the choice of frame. Using the formula for the Riemann curvature $R^1_{122}=0$, we can express it in terms of the director gradients as
\begin{equation} \label{eq:gaussian_curvature}
  K_\xi = \frac{s^2}{4} + \frac{q^2}{4} - \frac{1}{2} |\Delta|^2 + q \tau .
\end{equation}
We remark that the symmetric part of the shape operator $\nabla_{\perp}{\bf n}$, {\sl i.e.} $\frac{s}{2} I + \Delta$, has eigenvalues $\frac{s}{2} \pm |\Delta|/\sqrt{2}$ that we may view as the principal curvatures of the orthogonal plane field $\xi$~\cite{machon16}. (When the director is normal to a family of surfaces, {\sl i.e.} $q=0$, they are precisely the principal curvatures of the surfaces.) Hence, we may view this expression for the `Gaussian curvature' of $\xi$ as $K_{\xi} = \kappa_1 \kappa_2 + q^2/4 + q \tau$, where $\kappa_1$, $\kappa_2$ are the `principal curvatures'.

Finally, we remark that the description of the gradients in terms of structure functions leads to an alternative derivation of the Frank free energy. The choice of frame is arbitrary and not limited to the Frenet-Serret frame we have used so far; any other frame is related by a rotation ${\bf \bar{e}}_1 = \cos \theta{\bf e}_1 + \sin \theta{\bf e}_2, {\bf \bar{e}}_2 = {\bf n} \times {\bf \bar{e}}_1$, depending on an angle $\theta$. Under this change, the components $\omega^1_3, \omega^2_3$ of the connection 1-form transform as vector, giving $\bar{\omega}^1_3 = \cos \theta \omega^1_3 +\sin \theta \omega^2_3$, and similarly for $\bar{\omega}^2_3$, while $\bar{\omega}_1^2 = \omega_1^2 + d\theta$. These changes can also be expressed as changes in the structure functions. The energy may contain only those combinations of structure functions that are invariant under such changes, and also under reversing the sign of the director. Up to quadratic order, this gives those combinations of structure functions that result in the terms $q, s^2, q^2, |{\bf b}|^2,$ and $|\Delta|^2$. We may replace $|\Delta|^2$ using~\eqref{eq:saddle_splay} to arrive at the familiar form of the Frank free energy.

\section{Geometric Reconstruction in Three Dimensions}
\label{sec:geometric_reconstruction_3d}

Now we turn to the reconstruction problem in three dimensions. This is more challenging than in two dimensions due to the freedom in choosing the basis for the planes orthogonal to the director, and the more complicated relationship between the structure functions and the director gradients. To reconstruct a frame, it is necessary and sufficient to have all nine of the structure functions of that frame~\cite{cartan45,olver}; this means specifying more information than is contained purely in the gradient tensor $\nabla {\bf n}$, and indeed this appears to be unavoidable. To get around this constraint we may either specify additional information about the gradients of bend, specify vector quantities in addition to the scalar ones, or make additional assumptions about the geometry of the frame. We overview all these reconstruction methods below.

The most direct method of reconstructing the director uses the set $c_{ij}^k$ of nine structure functions along with the nine algebraic and geometric compatibility conditions. The unknown frame ${\bf e}_1, {\bf e}_2, {\bf e}_3 = {\bf n}$ with structure functions $c_{ij}^k$ can be expressed in terms of the coordinate frame ${\bf e}_{x_j}$ using a rotation matrix $R$, so that ${\bf e}_i = \sum_j R_{ij}{\bf e}_{x_j}$. To reconstruct the director, it suffices to find the entries of $R$. The compatibility conditions give us a set of nine linear equations for the entries of $R$ in terms of $c_{ij}^k$ and the tensor  $D_{ijl}^k = \nabla_{{\bf e}_{x_l}} c_{ij}^k$, where the gradient is taken along the coordinate direction ${\bf e}_{x_l}$.

We rewrite the algebraic compatibility conditions as
\begin{equation}
  \sum_{i < j} \sum_{l} \epsilon_{ijl} \left(\nabla_{{\bf e}_l} c_{ij}^k + c_{ij}^k s_l\right) = 0,
\end{equation}
for each $k=1,2,3$, where $\epsilon_{ijl}$ is the Levi-Civita tensor, and for compactness we have written the divergence of the vector field ${\bf e}_k$ as $s_k = \sum_i c_{ik}^i$. Expressing the gradient along ${\bf e}_l$ in terms of $R$ and the gradients along the coordinate directions, we find,
\begin{equation} \label{eq:Algebraic}
  \sum_{i<j} \sum_{l,m} R_{lm} \,\epsilon_{ijl} D_{ijm}^k = A_k,
\end{equation}
where we have defined $A_k = -\sum_{i<j}\sum_{l} \epsilon_{ijl} c_{ij}^k s_l$. For the geometric compatibility conditions, we first define $B_{ijk}^l = 2\sum_m c_{ijm} \omega_{mk}^l - \omega_{jk}^m \omega_{im}^l + \omega_{ik}^m \omega_{jm}^l$. Then the vanishing of the Riemann tensor implies, through~\eqref{eq:Riemann},
\begin{equation} \label{eq:Geometric}
  \sum_m R_{im} \bigl( D_{jkl}^m + D_{ljk}^m + D_{lkj}^m \bigr) - R_{jm} \bigl( D_{ikl}^m + D_{lik}^m + D_{lki}^m \bigr) = B_{ijk}^l .
\end{equation}
Together, Eqs.~\eqref{eq:Algebraic} and~\eqref{eq:Geometric} give a set of nine linear equations for the nine functions $R_{ij}$. To reconstruct the frame, we invert this system of linear equations to get an expression for $R_{ij}$ in terms of the known quantities $D_{ijk}^l$, $A_k$, and $B_{ijk}^l$.

Observe that each structure function $c_{ij}^k$ and each component of $D$ appears in these equations, so we cannot eliminate any of these quantities. As we have remarked, $c_{12}^1, c_{12}^2$ and their gradients cannot be obtained solely from $\nabla {\bf n}$. We can obtain them from the divergence, twist, and bend of the bend vector field ${\bf b}$, which are higher-order gradients of the director. As quadratic functions of the director gradients, they do not appear in the Oseen--Frank energy, but would appear in a higher-order energy functional. Energy functionals that have been proposed for twist-bend nematic materials make use of an additional direction, the polarisation, whose gradients appear in the energy~\cite{selinger2013,binysh2020}; in such materials the gradients of polarisation would also serve to provide the required structure functions.

This method reconstructs the entire frame, rather than just the director. A second method that reconstructs only the director itself mimics the approach in two dimensions by using the compatibility conditions to find the projection of the director onto a pair of distinct vector fields. If the structure functions are specified for the Frenet--Serret frame, then the sectional curvature $K_{23}$ gives the projection of the director on the vector $\nabla\left(\tfrac{s}{2}-\Delta_1 \right)$, Eq.~\eqref{eq:k23}. The component $R_{133}^2$ of the Riemann curvature tensor gives the projection on $\nabla\left(\tfrac{q}{2}+\Delta_2 \right)$. Assuming these directions are not colinear, we can use them to reconstruct the director. Set ${{\bf \bar{e}}}_1 = \tfrac{\nabla (s/2-\Delta_1)}{|\nabla (s/2-\Delta_1)|}$ and ${{\bf \bar{e}}}_2 = \tfrac{\nabla (q/2+\Delta_2)}{|\nabla (q/2+\Delta_2)|}$. Since we know the projection of the director on these unit vectors, we can compute the component in the orthogonal direction ${{\bf \bar{e}}}_1 \times {{\bf \bar{e}}}_2$, and hence reconstruct the director. In this approach we still need the structure functions $c_{12}^1$ and $c_{12}^2$.

Alternatively, the formula~\eqref{eq:saddle_splay} gives the projection of the director onto the vector $\nabla s$
\begin{equation}
  {\bf n} \cdot \nabla s = \nabla \cdot {\bf b} + \frac{q^2}{2}-\frac{s^2}{2} - |\Delta|^2,
\end{equation}
while~\eqref{eq:projector_q} gives the projection onto $\nabla q$
\begin{equation}
  {\bf n} \cdot \nabla q = {\bf n} \cdot \nabla \times {\bf b} - qs.
\end{equation}
This requires us to know the scalar quantities $s,q,|\Delta|$, ${\bf n} \cdot \nabla \times {\bf b}$ and $\nabla \cdot {\bf b}$; again, the latter two are not determined by $\nabla {\bf n}$ alone. Again, we can reconstruct the director from this information provided that $\nabla s$ and $\nabla q$ are not zero and are nowhere colinear.

These two methods require only scalar quantities but if vector data is given then other methods are possible. For example, if we specify a unit vector $\bar{{\bf e}}_2$ with the promise that the director is everywhere orthogonal to it, then we can reduce to the two-dimensional case and reconstruct the director from its splay $s$ and curvature $\kappa$. To do this, we compute the Gaussian curvature $K$ of the planes orthogonal to $\bar{{\bf e}}_2$, and introduce the rotation $J = \bar{{\bf e}}_2 \times$. Then Eq.~\eqref{eq:compatibility_2d_2} holds, and the director is given by Eq.~\eqref{eq:reconstruction_2d} with the appropriate $K$ and $J$. In particular, if we are given the splay and the bend as a vector, then this is sufficient information to reconstruct the director.

Finally, we describe an approach to the reconstruction problem based on ideas from contact topology~\cite{geiges08}, which has been employed recently in the study of cholesteric liquid crystals~\cite{machon17,pollard19}. For this approach we make use of the field of planes $\xi$ orthogonal to the director. In contact topology, plane fields are studied by the characteristic foliation they induce on a surface $S$: this is simply the intersection of the planes $\xi$ with $S$. The characteristic foliation determines the topology of the director in a neighbourhood of the surface; with knowledge of the gradients of the director we can get its geometry right as well.

To reconstruct the director fully we need not just one surface, but a collection of surfaces which span the region on which we want to reconstruct the director. We will use the plane surfaces $S_z$ where the value of the $z$-coordinate is constant. Other families of surfaces can also be used and may be helpful for studying other geometric features depending on the symmetries of the director.

\begin{figure}[tb]
\centering
\includegraphics[width=0.5\textwidth]{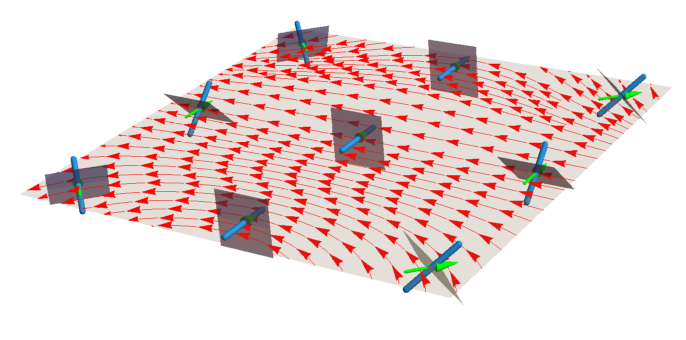}
\caption{Along a surface (grey) the director (blue) determines two orthogonal vector fields that give a frame on the surface: its projection into the surface ${\bf n}_\perp$ (green) and the characteristic foliation ${\bf \bar{e}}_1$ (red streamlines). The latter is the rotation of  ${\bf n}_\perp$ ninety degrees anticlockwise around the surface normal, but can also been seen as the vector field that directs the intersection of the planes $\xi$ orthogonal to ${\bf n}$ (gray) with the surface.}
\label{fig:fig2}
\end{figure}

Let ${\bf n}_\perp$ denote the unit vector along the projection of the director into the surface $S_z$. The characteristic foliation is directed by a unit vector field ${\bf \bar{e}}_1$ that is orthogonal to ${\bf n}_{\perp}$; see Fig~\ref{fig:fig2}. As ${\bf \bar{e}}_1$ is orthogonal to the director we can expand it to a frame by setting ${\bf \bar{e}}_2 = {\bf n} \times {\bf \bar{e}}_1$. This frame is highly non-generic and hence we can expect less information is required to reconstruct it than in the general case. Denote its structure functions by $\bar{c}_{ij}^k$.

The vector fields ${\bf \bar{e}}_1, {\bf n}_\perp$ are an orthonormal frame for the surface $S_z$, and hence can written as ${\bf n}_\perp = \cos(\psi){\bf e}_x + \sin(\psi){\bf e}_y$ and ${\bf \bar{e}}_1 = -\sin(\psi){\bf e}_x + \cos(\psi){\bf e}_y$, for some function $\psi$. We compute that ${\bf \bar{e}}_1 \cdot \nabla \times {\bf \bar{e}}_1 = -\partial_{z}\psi$. In terms of the structure functions of the frame the twist of ${\bf \bar{e}}_1$ is $-\bar{c}_{23}^1$, so $\psi$ satisfies the differential equation $\partial_z\psi = \bar{c}_{23}^1$. Presuming the structure function given, solving this differential equation only determines $\psi$ up to a function $\psi_0(x,y)$ of $x$ and $y$ alone, which we can compute if we have fixed the charcteristic foliation, or ${\bf n}_{\perp}$, along $S_0$.

This determines the direction of the projection of the director into each surface $S_z$, so that it remains only to determine the $z$-component. Write ${\bf n} = \cos(\phi) {\bf e}_z + \sin(\phi){\bf n}_\perp$ and ${\bf \bar{e}}_2 = {\bf n} \times {\bf \bar{e}}_1 = \cos(\phi) {\bf n}_\perp - \sin(\phi){\bf e}_z$ for an angle $\phi$ to be found. To relate $\phi$ to the structure functions, we compute the Lie brackets between elements of this frame. Write $[{\bf n}_\perp, {\bf \bar{e}}_1] = f_1 {\bf n}_\perp + f_2 {\bf \bar{e}}_1$ for the Lie bracket between these vector fields on the surface $S_z$. Note that this bracket has no component in the $z$-direction as ${\bf n}_\perp, {\bf \bar{e}}_1$ span integral surfaces. We compute that $[{\bf n}_\perp, {\bf \bar{e}}_1] = -\partial_x \psi \,{\bf e}_x - \partial_y \psi \,{\bf e}_y$, which leads to
\begin{equation}
 \begin{aligned}
   f_1 &= - \cos(\psi)\partial_x \psi - \sin(\psi)\partial_y \psi, \\
   f_2 &= \sin(\psi)\partial_x \psi - \cos(\psi)\partial_y \psi.
 \end{aligned}
\end{equation}
The interpretation of $f_1,f_2$ is the same as for structure functions in two dimensions, as explained in \S\ref{sec:geometric_reconstruction_2d}: $-f_1$ is the curvature of the integral curves of ${\bf n}_\perp$, while $-f_2$ is the divergence of ${\bf n}_\perp$. It is worth noting that if we wished to instead prescibe $f_1$ and $f_2$, we could use this to reconstruct ${\bf n}_\perp$ without using the structure function $\bar{c}_{23}^1$ by applying the method of Niv \& Efrati for two-dimensional reconstruction outlined in \S\ref{sec:geometric_reconstruction_2d}.

Next we compute $[{\bf \bar{e}}_1, {\bf n}]$. To do this we use $[{\bf n}_\perp, {\bf \bar{e}}_1]$ and two other brackets, $[{\bf n}_\perp, {\bf e}_z] = -\partial_z\psi \,{\bf \bar{e}}_1$ and $[{\bf \bar{e}}_1, {\bf e}_z] = \partial_z\psi \,{\bf n}_\perp$. From these we determine that $\bar{c}_{13}^1 = -f_2\sin(\phi)$. Consequently, whenever $f_2 \neq 0$ we can write
\begin{equation} \label{eq:reconstruction_formula}
 {\bf n} = \biggl[ 1-\biggl( \frac{\bar{c}_{13}^1}{f_2} \biggr)^2 \biggr]^{1/2} \,{\bf e}_z - \frac{\bar{c}_{13}^1}{f_2} \bigl( \cos(\psi) \,{\bf e}_x + \sin(\psi) \,{\bf e}_y \bigr) .
\end{equation}
There are two choices for the sign of the square root: only one sign will yield the correct director. We then have a geometric reconstruction formula similar to the two-dimensional case.

\section{Director Distortions and Lie Groups} \label{sec:lie_groups}
In this section we will examine a few classes of directors -- namely those with pure distortions, uniform distortions, and quasi-uniform distortions -- from the perspective of Lie algebras and groups. There is some overlap between the results of this section and a recent paper of Sadoc~\textit{et al.}~\cite{sadoc20}, however the Lie group perspective we adopt here gives a novel approach to the problem.

\subsection{Curvature of Pure Distortions}

When the structure functions of a frame ${\bf e}_1, {\bf e}_2, {\bf e}_3 = {\bf n}$ are constant, they describe a Lie algebra. It is a well-known fact that every Lie algebra uniquely determines a homogeneous space, a Lie group~\cite{cartan04}. Such groups are abstract spaces where the given structure functions could be realised, even if they cannot not be realised in Euclidean space. The sectional curvatures $K_{ij}$ of these Lie groups give insight into what spatial curvature is required to realise particular special distortion modes, while the group itself describes the natural symmetries associated to that director distortion. The sectional curvatures can be computed in terms of the structure functions, or the components of the connection 1-form, using~\eqref{eq:Riemann} and are explicitly given by
\begin{equation} \label{eq:Lie_curvature}
  \begin{split}
    K_{ij} = \ &\nabla_{{\bf e}_i} c_{ij}^{j} - \nabla_{{\bf e}_j} c_{ij}^{i} - \sum_k \frac{1}{2}c_{ij}^{k} \bigl( c_{ij}^{k} + c_{kj}^{i} - c_{ki}^{j} \bigr) \\
              &- \frac{1}{4}\bigl( (c_{ij}^{k})^2 - (c_{jk}^{i}-c_{ki}^{j})^2 \bigr) - c_{ki}^{i}c_{kj}^{j} .
  \end{split}
\end{equation}
This expression was derived by Milnor~\cite{milnor76} in his analysis of the curvature of metrics on Lie groups, although he only considers the case where the structure functions are constant and thus his formulae do not contain the gradient terms present in~\eqref{eq:Lie_curvature}. We use the formula to understand the curvature associated to pure splay, twist, bend, or anisotropic orthogonal gradient distortions. Moreover, certain Lie groups admit a projection into flat Euclidean space, allowing us to realise the idealised texture as a frustrated texture in Euclidean space.

Firstly, consider a director with constant nonzero twist $q$, but vanishing splay, bend, and $\Delta$. This requires the structure functions $c_{12}^3 = -q$, $c_{23}^1=c_{31}^2$, with $c_{12}^1,c_{12}^2$ left undetermined and all other structure functions vanishing. A minimal model for this has $c_{12}^1=c_{12}^2=0$ and either $c_{23}^1 = -q$ or $+q$. The former choice yields the Lie algebra of $SU(2)$, the 3-sphere, and the set of vector fields described previously by Sethna~\textit{et al.}~\cite{sethna1983} which, when projected into Euclidean space, give the familiar double-twist director. The latter choice gives the Lie algebra of the group $SL(2, \mathbb{R})$, which consists of real, traceless $2 \times 2$ matrices. The group can be identified with the anti-de Sitter space-time $AdS_3$, that is, the unit sphere in $\mathbb{R}^4$ equipped with the pseudometric with signature $(1,1,-1,-1)$. This texture is an alternative realisation of `double-twist' where the twisting in the orthogonal directions has the opposite sense of handedness to the twisting in the director: see Sadoc~\textit{et al.}~\cite{sadoc20} for a further discussion of this example.

One may also take $c_{23}^1=c_{31}^2 = 0$ for a `local twist' director, which yields the Lie algebra of the Heisenberg group. This group has a natural coordinate parameterisation $x,y,z$ in which the director is given by
\begin{equation}
  {\bf m} = {\bf e}_z + x \,{\bf e}_y - y \,{\bf e}_x,
\end{equation}
which, after normalisation in the Euclidean metric, ${\bf n} = {\bf m}/|{\bf m}|$, is the standard example of local twist, see for example Selinger~\cite{selinger18}. This vector field is also recognisable as the Darboux normal form of the standard tight contact structure in contact topology~\cite{geiges08}.

We remark that for the Frenet-Serret framing that we predominantly adopt $c_{23}^1=\tau-q/2$, so that these three examples differ by the (constant) value of the torsion, $\tau = -q/2$, $3q/2$ and $q/2$, respectively. These specific values are not crucially important as the relevant distinction between the Lie algebras comes from the sign, or vanishing, of $c_{23}^1$, whose magnitude can be made to match that of $c_{12}^3$ by an anisotropic scaling. The case of `pure twist', with vanishing torsion, falls in the $SU(2)$ class.

We make a final observation about pure twist: Sadoc~\textit{et al.} conjucture that a positive Ricci curvature along one direction is a necessary condition for a pure nonzero twist state~\cite{sadoc20}. This conjecture is correct and follows immediately from the computation of $K_{13}+K_{23}$~\eqref{eq:saddle_splay}, which in a pure twist case is simply equal to $q^2/2$, so that the Ricci curvature along the director is always positive.

Next consider pure splay and pure bend. Pure splay requires $c_{13}^1=c_{23}^2 = s/2$, and we may take all other structure functions equal to zero for a minimal model. The resulting Lie group is consequently a space of constant negative curvature~\cite{milnor76}. Even if we allow ourselves maximum freedom, with $c_{12}^1,c_{12}^2$ free and $c_{23}^1=c_{31}^2$ to ensure the vanishing of $\Delta$, then all sectional curvatures will still be negative. Pure bend requires $c_{13}^3, c_{23}^3$ nonzero. If we take $c_{12}^1 = c_{23}^3$ and $c_{12}^2 = -c_{13}^3$, and similarly we find the curvature is negative. We note that in two dimensions it is also true that constant, nonzero pure bend or pure splay states can only occur on a negatively-curved space, as follows from~\eqref{eq:compatibility_2d}. Both pure splay and pure bend correspond to the Lie group of type V in the Bianchi classification when $c_{23}^1=c_{31}^2=0$, and type $\text{VII}_a$ when these structure functions are nonzero.

Finally, we ask for the possibility of pure $\Delta$. We can assume without loss of generality that $\Delta$ is off-diagonalised, so this mode of distortion requires $c_{23}^1=-c_{31}^2$ and $c_{13}^1=c_{23}^2=0$. Setting all other structure functions equal to zero gives a space where the curvature of planes orthogonal to the director is positive, whereas the curvature of planes containing the director is negative, suggesting this is the natural form of curvature associated to this distortion. This pure distortion corresponds to the unimodal Lie group $E(1,1)$, the isometries of Minkowski 2-space~\cite{milnor76}.

\subsection{Uniform Euclidean Distortions}
\label{subsec:uniform}

One may ask when it is possible to have a director with uniform distortions in ordinary Euclidean space. It is known that all such directors are given by the heliconical director~\cite{virga19},
\begin{equation}
  {\bf n} = \cos(\theta) \,{\bf e}_z + \sin(\theta) \bigl( \cos(qz) \,{\bf e}_x \pm \sin(qz) \,{\bf e}_y \bigr),
\end{equation}
for some choice of $\theta, q$ constant. This encapsulates both the cholesteric ground state ($\theta = \pi/2$) and the nematic ground state ($\theta = 0$) as limits.

We provide an alternative argument for this classification based on the framework we have adopted in this paper. A uniform director in Euclidean space has constant structure functions. Using the compatibility conditions, we can examine which sets of constant $c_{ij}^k$ are permissable. Any set of structure functions satisfying the algebraic compatibility conditions defines a Lie algebra, and conversely every Lie algebra gives rise to a frame satisfying the algebraic compatibility conditions. Satisfying the geometric compatibility conditions as well is then equivalent to the Lie group associated to the Lie algebra having a flat metric. This Lie group, taken along with rotations and about the director ${\bf n}$ and the nematic symmetry ${\bf n} \mapsto -{\bf n}$, can be seen as the symmetry group of the texture described by ${\bf n}$.

Three-dimensional Lie algebras have been classified up to isomorphism~\cite{patera76}. Up to this equivalence, the only Lie algebras giving rise to flat Lie groups are the trivial algebra with all Lie brackets being equal to zero, which corresponds to the coordinate basis of $\mathbb{R}^3$, and the algebra defined by the brackets $[{\bf e}_1, {\bf e}_2] = 0$, $[{\bf e}_3, {\bf e}_1] = {\bf e}_2$ and $[{\bf e}_2, {\bf e}_3] = {\bf e}_1$, which is the Lie algebra of the Euclidean group $E(2)$, the group of rigid motions of the Euclidean plane. A concrete realisation of the latter as a set of vector fields in $\mathbb{R}^3$ is given by
\begin{equation} \label{eq:chol_frame}
  \begin{aligned}
    {\bf e}_1 &= \cos(z) \,{\bf e}_x + \sin(z) \,{\bf e}_y, \\
    {\bf e}_2 &= - \sin(z) \,{\bf e}_x + \cos(z) \,{\bf e}_y, \\
    {\bf e}_3 &= {\bf e}_z.
  \end{aligned}
\end{equation}
Any combination of these three vector fields with constant coefficients, {\sl e.g.} the heliconical director, or those involving rescalings of the coordinate directions, {\sl e.g.} $z \mapsto qz$, will also give a distortion frame with constant structure functions. The classification of Lie algebras implies that, up to a coordinate parameterisation, these are the only possible cases.

\subsection{Quasi-Uniform Euclidean Distortions}

Pedrini \& Virga~\cite{virga20} have defined a notion of quasi-uniform distortions as those that are in constant proportion to one another. In our language, this means the structure functions are proportional to constants with common proportionality, $c_{ij}^k = f a_{ij}^k$ for $f$ a function and $a_{ij}^k$ constant. One can construct examples by assuming that $a_{ij}^k$ are the structure constants of a Lie algebra. Let ${\bf e}_i$ be the distortion frame defined in \S\ref{sec:geometric_reconstruction_3d}, and suppose that $c_{ij}^k$ are the structure functions of this frame. Denote by $K_{ij}(a)$ the sectional curvatures of the Lie group whose Lie algebra has structure constants $a_{ij}^k$. The algebraic compatibility conditions for $c_{ij}^k$ then reduce to a set of differential equations that, along with the vanishing of the Riemann tensor or the sectional curvatures, serve as compatibility conditions for the function $f$,
\begin{equation} \label{eq:compat_f}
  \begin{aligned}
    a_{23}^k\nabla_{{\bf e}_1} f + a_{31}^k\nabla_{{\bf e}_2} f + a_{12}^k \nabla_{{\bf e}_3} f = 0, \\
    K_{ij}(a) + a_{ij}^j \nabla_{{\bf e}_i}f  + a_{ji}^i \nabla_{{\bf e}_j} f = 0,
  \end{aligned}
\end{equation}
for $i,j,k$ running from 1 to 3. One may then determine those choices of Lie algebra for which this equation has solutions, which give necessary and sufficient conditions for the existence of a frame with these structure functions. For example, consider the Lie algebra with the only nonzero structure constants $a_{23}^1=a_{31}^2 =1$ defined by the vector fields~\eqref{eq:chol_frame}. The compatibility conditions~\eqref{eq:compat_f} for the function $f$ then reduce to $\partial_x f = \partial_y f = 0$, and therefore we may choose $f$ to be any function of $z$ alone. The frame giving rise to the structure functions $c_{ij}^k$ is:
\begin{equation} \label{eq:chol_frame2}
  \begin{aligned}
    {\bf e}_1 &= \cos g(z) \,{\bf e}_x + \sin g(z) \,{\bf e}_y, \\
    {\bf e}_2 &= - \sin g(z) \,{\bf e}_x + \cos g(z) \,{\bf e}_y, \\
    {\bf e}_3 &= {\bf e}_z,
  \end{aligned}
\end{equation}
where $\partial_z g = f$. As before, any unit director which is a constant combination of these three vector fields will be quasi-uniform, {\sl e.g.} a quasi-uniform heliconical director ${\bf n} = \cos(\theta)\,{\bf e}_z + \sin(\theta)(\cos g(z) \,{\bf e}_x \pm \sin g(z) \,{\bf e}_y)$ with $\theta$ constant. These then give examples of quasi-uniform directors with twist and bend distortions that vary in the $z$ direction according to (the antiderivative of) $f$, but still have vanishing splay.

More generally, if the Lie algebra can be reduced to a form where the structure constants $a_{ij}^{j}$ vanish, then we see that it is impossible for there to be a quasi-uniform director corresponding to this Lie algebra unless the Lie algebra itself is flat. These algebras are the unimodular Lie algebras~\cite{milnor76}, and the two flat unimodular algebras are, as we have already noted, $\mathbb{R}^3$ and the Euclidean group $E(2)$ which determine the nematic and the cholesteric/heliconical states respectively. Any quasi-uniform director corresponding to the latter is defined by a taking a constant combination of the vectors in the frame~\eqref{eq:chol_frame2}.

When a Lie group is not unimodal, it has a basis ${\bf e}_j$ such that
\begin{equation}
  \begin{aligned}
    {[{\bf e}_1, {\bf e}_3]} &= A{\bf e}_1 + B {\bf e}_2, \\
    [{\bf e}_2, {\bf e}_3] &= C{\bf e}_1 + D {\bf e}_2,
  \end{aligned}
\end{equation}
for constants $A,B,C,D$ such that $A+D=2$~\cite{milnor76}. The second set of conditions in (\ref{eq:compat_f}) then imply that $K_{12}(a) = 0$, which is satisfied when $B=-C$, and further that $A(\nabla_3 f - A) = D(\nabla_3 f - D) = 0$, which implies that either one of $A, D$ vanishes, or that $A=D=1$, and that $\nabla_3 f$ is constant and nonzero. Moreover, the first set of conditions imply that $f$ is constant along ${\bf e}_1, {\bf e}_2$. Since the curl of the ${\bf e}_3$ direction vanishes, we can assume that it is the gradient of a function $h$. The vector fields ${\bf e}_1, {\bf e}_2$ must be a pair of orthogonal vector fields orthogonal to ${\nabla h}$. As $[{\bf e}_1, {\bf e}_2]$ vanishes the level sets of $h$ must be flat, so we may take ${\bf e}_3 = {\bf e}_z$, and consequently these cases all reduce to the case of a uniform director.

Thus we have shown that, when the $a_{ij}^{k}$ are chosen to be the structure constants of a Lie algebra, then the only possible quasi-uniform directors are those in the family (\ref{eq:chol_frame2}).

More options are possible if we allow for singularities in the director, or for the function $f$ to have singular behaviour. For instance, the director ${\bf n} = \cos(\phi){\bf e}_z + \sin(\phi){\bf e}_r$ in cylindrical coordinates $(r,\theta,z)$, which has ${\bf n}_\perp = {\bf e}_r$ and ${\bf e}_1 = r{\bf e}_\theta$, is a quasi-uniform pure-splay texture provided that $0 \leq \phi \leq \pi/2$ is constant, in which case we have structure functions $c_{12}^1 = -\frac{1}{r}\cos(\phi), c_{13}^1 = \frac{1}{r}\sin(\phi)$, with the others vanishing. The function $f$ here is $1/r$, which becomes undefined along the $z$ axis. Quasi-uniform pure-bend states can be defined similarly, by ${\bf n} = \cos(\phi){\bf e}_z + \sin(\phi){\bf e}_\theta$, for $0 \leq \phi \leq \pi/2$ constant.

\section{Discussion}
\label{sec:discussion}
In this paper we have given the connection between the distortions of a liquid crystal and the metric connection, which allows us to derive a set of compatibility conditions between those distortions. We describe several different methods for reconstructing a director from its gradients, which can be applied to a variety of different geometric problems. Interestingly, these reconstruction formulas show that it is not possible to reconstruct a director from the gradient tensor $\nabla {\bf n}$ alone: we either require second order gradients of the director such as the gradients of the bend vector field, or we must have the gradient tensor expressed with respect to a particular frame that reduces the degrees of freedom.

Our approach connects the distortion frame of a director to the theory of Lie groups, which give natural examples of curved spaces in which one may realise directors with only a single distortion mode nonzero. This allows us to understand how each individual distortion mode couples to curvature, and well as elucidating the symmetries associated with the distortion mode generalising earlier work that showed the positively-curved 3-sphere was the natural space for the double-twist distortions common in cholesteric materials. This approach suggests a programme of describing directors based on their local symmetry groups and the way the symmetry group changes as we move in space. This will be addressed in future work.

\section*{Acknowledgements}
This work is supported by the UK ESPRC through Grant No. EP/L015374/1.


\end{document}